\documentclass[12pt]{article}
\usepackage{amssymb,amsmath,graphicx,color}
\newcommand{\be}{\begin{equation}}
\newcommand{\ee}{\end{equation}}
\newcommand{\bear}{\begin{eqnarray}}
\newcommand{\eear}{\end{eqnarray}}
\newcommand{\beal}{\begin{align}}
\newcommand{\eeal}{\end{align}}
\newcommand{\ba}{\begin{array}}
\newcommand{\ea}{\end{array}}
\newcommand{\nn}{\nonumber}

\begin{document}

\vspace{9mm}

\begin{center}
{{{\Large \bf Mass-Deformed Super Yang-Mills Theories from M2-Branes
with Flux}
}\\[17mm]
Yoonbai Kim$^{1}$,~~O-Kab Kwon$^{1}$,~~D.~D. Tolla$^{1,2}$\\[3mm]
{\it $^{1}$Department of Physics,~BK21 Physics Research Division,
~Institute of Basic Science,\\
$^{2}$University College,\\
Sungkyunkwan University, Suwon 440-746, Korea}\\[2mm]
{\tt yoonbai@skku.edu,~okab@skku.edu,~ddtolla@skku.edu} }
\end{center}

\vspace{20mm}

\begin{abstract}
We consider $(2+1)$-dimensional mass-deformed SYM theories and their
M-theory origin. These are obtained from MP Higgsing of ABJM theory
with constant flux and fixing the mass terms via supersymmetry
completion. Depending on the choice of the flux, we obtain ${\cal
N}=1$, ${\cal N}=2$, and ${\cal N}=4$ mass-deformed SYM theories.
For each of these cases we solve the vacuum equation and obtain the
fuzzy two ellipsoid solution for the first two cases. We also
discuss the D-brane interpretation of the obtained mass-deformed SYM
theories.

\end{abstract}

\newpage

\tableofcontents

\section{Introduction}

Proposal of the theories describing the low energy dynamics of
multiple M2-branes has drastically improved our understanding of
M-theory ~\cite{Bagger:2006sk,Gustavsson:2007vu,Aharony:2008ug}.
Subsequently, various deformations of Bagger-Lambert, Gustavsson
(BLG) and Aharony-Bergman-Jafferis-Maldacena (ABJM) theories have
been discussed, including maximal supersymmetry preserving
mass-deformation of BLG theory~\cite{Gomis:2008cv,Hosomichi:2008qk}
and ABJM theory~\cite{Hosomichi:2008jb,Gomis:2008vc}, higher
derivative corrections to BLG theory~\cite{Ezhuthachan:2009sr},
addition of matter multiplets in fundamental representations in ABJM
theory~\cite{Gaiotto:2009tk,Hikida:2009tp}, introduction of the
Wess-Zumino (WZ) type couplings to the background form fields
\cite{Li:2008eza,Ganjali:2009kt,Kim:2009nc,Lambert:2009qw,Sasaki:2009ij,KKNT,Allen:2011pm},
and so on. One noteworthy aspect is the fact that the dimensional
reduction of the theory of multiple M2-branes via the
Mukhi-Papageorgakis (MP) Higgsing procedure~\cite{Mukhi:2008ux}
provides the description of low energy dynamics of multiple
D2-branes, the ${\cal N}=8$ super Yang-Mills (SYM) theory in
$(2+1)$-dimensions.  Since the MP Higgsing of the undeformed
theories gives the (2+1)-dimensional SYM theory, it is intriguing to
investigate which of the deformations allow the Higgsing procedure
for dimensional reduction and what are their resultant theories.

In this paper we are interested in the supersymmetry preserving
mass-deformations in the ABJM theory, which are generated by turning
on a transverse constant four-form field strength and the dual
seven-form field strength~\cite{Kim:2009nc,KKNT,Lambert:2009qw}.
Though it is natural to begin with the maximal supersymmetry
preserving mass-deformation, the presence of quadratic mass terms
for all the transverse scalar fields makes the bosonic potential
have no flat direction. Since the MP Higgsing procedure is not
applicable to this case, we shall start with the ABJM theory
deformed by a WZ-type coupling to a six-form gauge field with an
arbitrary constant seven-form field strength. In general this
deformation breaks supersymmetry, however, the MP Higgsing procedure
is applicable for this model and the results is a (2+1)-dimensional
Yang-Mills matter Lagrangian involving the Myers coupling to
five-form gauge field~\cite{Myers:1999ps}.  From the side of type
IIA string theory, we may have some mass-deformed (2+1)-dimensional
SYM theories which can be obtained through a circle compactification
of one world-volume direction in the Polchinski-Strasller ${\cal
N}=1^*$ and ${\cal N}=2^*$ theories in
$(3+1)$-dimensions~\cite{Polchinski:2000uf}. Keeping these two sets
of (2+1)-dimensional Yang-Mills theories in mind, one may expect to
preserve some supersymmetry for the former theory and reproduce an
equivalent of the later theory. We show that preservation of
supersymmetries will fix the values of the nonvanishing components
of the five-form gauge field as well as the mass terms for the
fermionic and the scalar fields. Depending on the choice of the
flux, we obtain ${\cal N}=2$ and ${\cal N}=4$ mass-deformed SYM
theories, which are linked with the ${\cal N}=1^*$ and ${\cal
N}=2^*$ theories, as well as a ${\cal N}=1$ theory. After fixing the
form of the five-form gauge field using supersymmetry invariance in
type IIA string theory, we combined this result with that of the MP
Higgsing procedure to determine the form of the corresponding
six-form gauge field in M-theory. The resulting six-form gauge field
is different from the one which generates the maximal supersymmetry
preserving mass-deformation in ABJM theory.

The remaining part of the paper is organized as follows. In section
\ref{ABJM-H} we apply the MP Higgsing procedure to the original ABJM
theory and verify that this results in a supersymmetry enhancement
and reproduces the ${\cal N}=8$ SYM in (2+1)-dimensions. In section
\ref{fluxH} we apply the Higgsing procedure to the WZ-type coupling
for a constant seven-form field strength and obtain the
corresponding Myers coupling to a five form-field in type IIA string
theory. In section \ref{massSYM} we use the Myers coupling of the
type determined in section \ref{fluxH} and obtain the mass-deformed
SYM theories preserving ${\cal N}=1$, ${\cal N}=2$, or ${\cal N}=4$
supersymmetries. For each of these cases we will solve the vacuum
equation and obtain the fuzzy two ellipsoid solution in the first
two cases. We also discuss the D-brane interpretation of the
obtained mass-deformed SYM theories. In section \ref{Morg} we invert
the results of the Higgsing procedure and identify the possible
M-theory origin of the fluxes which generate supersymmetry
preserving mass-deformations in type IIA string theory. Section
\ref{conc} is devoted to conclusions and future research directions.

\section{Higgsing of the ABJM Theory}\label{ABJM-H}

Based on the BLG theory, Mukhi and Papageorgakis established a
Higgsing procedure~\cite{Mukhi:2008ux}, which reduces the theory
describing multiple M2-branes to theory of multiple D2-branes. An
extension to the ABJM theory with U($N$)$\times$U($N$) gauge group
was made in Ref.~\cite{Pang:2008hw} and they obtained
(2+1)-dimensional ${\cal N}=8$ SYM theory with U($N$) gauge group.
In their setup, the Higgsing procedure also gives a Lagrangian of a
free scalar field which is decoupled from the other fields. In this
section we recapitulate the calculation that the application of MP
Higgsing procedure to the same ABJM theory in the setup of
Ref.~\cite{KKNT} reproduce the same SYM theory without the decoupled
free scalar field.

The ABJM action~\cite{Aharony:2008ug} is given by a Chern-Simons matter theory with ${\cal N}=6$
supersymmetry and U($N$)$\times$U($N$) gauge symmetry,
\begin{align}\label{ABJMact}
S =\int d^3x\,\left({\cal L}_0 + {\cal L}_{{\rm CS}} + {\cal
L}_{{\rm ferm}} +{\cal L}_{{\rm bos}} \right),
\end{align}
where
\begin{align}
{\cal L}_0 &= {\rm tr}\left(-D_\mu Y_A^\dagger D^\mu Y^A +
i\Psi^{\dagger A} \gamma^\mu D_\mu \Psi_A\right),
\label{L0}\\
{\cal L}_{{\rm CS}} &= \frac{k}{4\pi}\,\epsilon^{\mu\nu\rho}\,{\rm
tr} \left(A_\mu \partial_\nu A_\rho +\frac{2i}{3}A_\mu A_\nu A_\rho
- \hat{A}_\mu \partial_\nu \hat{A}_\rho -\frac{2i}{3}\hat{A}_\mu
\hat{A}_\nu \hat{A}_\rho \right),
\label{LCS} \\
{\cal L}_{{\rm ferm}} &= -\frac{2\pi i}k{\rm tr}\Big( Y_A^\dagger
Y^A\Psi^{\dagger B}\Psi_B -Y^A Y_A^\dagger\Psi_B \Psi^{\dagger B}
+2Y^AY_B^\dagger\Psi_A\Psi^{\dagger B} -2Y_A^\dagger
Y^B\Psi^{\dagger A}\Psi_B
\nn \\
&\hskip 1.9cm  +\epsilon^{ABCD}Y^\dagger_A\Psi_BY^\dagger_C\Psi_D
-\epsilon_{ABCD}Y^A\Psi^{\dagger B}Y^C\Psi^{\dagger D} \Big),
\label{Lfe} \\
{\cal L}_{{\rm bos}} &=\frac{4\pi^2}{3k^2}{\rm tr}\Big(
Y^\dagger_AY^AY^\dagger_BY^BY^\dagger_CY^C
+Y^AY^\dagger_AY^BY^\dagger_BY^CY^\dagger_C
+4Y^\dagger_AY^BY^\dagger_CY^AY^\dagger_BY^C
\label{Lbo} \\
&\hskip 2cm -6Y^AY^\dagger_BY^BY^\dagger_AY^CY^\dagger_C \Big).\nn
\end{align}
The four complex
scalar fields $Y^A$ $(A=1,2,3,4)$ represent the eight directions $X^I$ $(I=1,\cdots,8)$
transverse to the M2-branes,
\begin{align}\label{YA}
Y^A=X^A+iX^{A+4}.
\end{align}

Let us take into account Higgsing of the bosonic sector of this
Lagrangian. For completeness we briefly summarize the Higgsing
procedure of the Ref.~\cite{KKNT} including the fermionic sector.
The first step is breaking the U($N$)$\times$U($N$) gauge symmetry
down to U($N$), so that the scalar fields are in the adjoint
representation of the unbroken U($N$). As a result, the transverse
scalars $X^I$ can be split into their trace and traceless part as
\begin{align}
&X^I={\check X}^I+i{\hat X}^{I}={\check X}_0^IT^0+i{\hat
X}_\alpha^{I}T^\alpha ,
\end{align}
where $T^0$ and $T^\alpha ~~ (\alpha=1,\cdots, N^2-1$) are the
generators of U(1) and SU($N$), respectively. Then the covariant
derivatives of the complex scalars become
\begin{align}
D_\mu Y^A=\tilde D_\mu X^A+i\tilde D_\mu
X^{A+4}+i\{A_\mu^-,X^A+iX^{A+4}\}, \label{cv2}
\end{align}
where $\tilde D_\mu X=\partial_\mu X+i[A_\mu^+,X]$ and
$A_{\mu}^\pm=\frac12(A_{\mu}\pm \hat A_{\mu})$.

The next step of the Higgsing procedure is to turn on vacuum
expectation value $v$ for the trace part of one of the complex
scalar fields,
\begin{align}\label{vev}
Y^A&=\frac v2 T^0\delta^{A4} +X^{A}+iX^{A+4}
\nn \\
&=\frac v2 T^0\delta^{A4} +\tilde X^{A}+i\tilde X^{A+4}.
\end{align}
In the second line we introduced eight Hermitian scalar fields in the adjoint
representation of the unbroken U($N$) as
\begin{align}\label{Xbar}
\tilde X^{A}={\check X}^{A}-{\hat X}^{A+4},\quad
 \tilde X^{A+4}={\check X}^{A+4}+{\hat X}^{A}.
\end{align}
Now we take a double scaling limit of large $v$ and large
Chern-Simons level $k$ with finite $v/k$.
To the leading order in $1/v$ the covariant derivatives \eqref{cv2} under
a gauge choice  $A_\mu^-\to A_\mu^--\frac 1v \tilde D_\mu ({\check
X}^8+{\hat X}^{4})$ become
\begin{align}
D_\mu Y^4&=\tilde D_\mu ({\check X}^4-{\hat X}^{8})+iv[A_\mu^-+\frac
1v(\tilde D_\mu ({\check X}^8+{\hat X}^{4}))] =\tilde D_\mu \tilde X^4+ivA_\mu^- ,
\nonumber\\
D_\mu Y^a&=\tilde D_\mu [({\check X}^a+i{\hat X}^{a})+i({\check
X}^{a+4}+i{\hat X}^{a+4})] =\tilde D_\mu \tilde X^{a}+i\tilde
D_\mu\tilde X^{a+4}, \qquad a=1,2,3.\label{covd}\end{align} Then the
gauge field $A_\mu^-$ is an auxiliary field in the resulting action
and can be integrated out. This completes the Higgsing of the
bosonic part of the ABJM theory, which results in the Lagrangian
\begin{align}\label{Lbos} {\cal L}_{\rm bos}=\frac 1{g^2}{\rm
tr}\Big(-\tilde D_\mu \tilde X^i \tilde D^\mu \tilde X^i-\frac12
\tilde F_{\mu\nu}\tilde F^{\mu\nu} + \frac12[\tilde X^i,\,\tilde
X^j]^2\Big),\quad i,j=1,\cdots,7,
\end{align}
where $\tilde F_{\mu\nu}$ is the field strength of the gauge field
$A_\mu^+$ and $g=\frac{2\pi v}k$ is the Yang-Mills coupling. In the
last step we have rescaled the scalar fields as $\tilde X^i\to
\frac1g\tilde X^i$.

In order to perform the Higgsing procedure in the fermionic sector
we start by splitting the complex fermionic fields as
\begin{align}\label{psiAA}
\Psi_A=\psi_A+i\psi_{A+4}.
\end{align}
Once the U($N$)$\times$U($N$) gauge symmetry is broken down to
U($N$), the fermions are also in the adjoint representation of the
unbroken U($N$). Then we can split the trace and the traceless parts
of $\psi_A$ and $\psi_{A+4}$,
\begin{align}
&\psi_A={\check \psi}_A+i{\hat \psi}_A=({\check
\psi}_A)_0T^0+i({\hat\psi}_A)_\alpha T^\alpha, \nonumber\\
&\psi_{A+4}={\check \psi}_{A+4}+i{\hat \psi}_{A+4}=({\check
\psi}_{A+4})_0T^0+i({\hat\psi}_{A+4})_\alpha T^\alpha.
\end{align}
 By introducing eight Hermitian fermionic fields in the adjoint representations,
\begin{align}\label{Psi}
\tilde\psi_A={\check\psi}_A-{\hat\psi}_{A+4},\quad\quad
 \tilde\psi_{A+4}={\check\psi}_{A+4}+{\hat\psi}_A,
\end{align}
we rewrite the fermions \eqref{psiAA} as
\begin{align}\label{PsiA}
\Psi_A={\check\psi}_A
+i{\hat\psi}_A+i({\check\psi}_{A+4}+i{\hat\psi}_{A+4})=
\tilde\psi_A+i\tilde\psi_{A+4}.
\end{align}
In the double scaling limit, the covariant derivatives of the
fermionic field to the leading order in $1/v$ become
\begin{align}
D_\mu \Psi_A=\tilde D_\mu
\Big[{\check\psi}_A+i{\hat\psi}_A+i({\check\psi}_{A+4}+i{\hat\psi}_{A+4})\Big]
=\tilde D_\mu \tilde\psi_A+i\tilde D_\mu\tilde\psi_{A+4}.
\label{covder}\end{align} Substituting \eqref{PsiA}--\eqref{covder}
into the ABJM Lagrangian and rescaling the fermions as
$\tilde\psi_{r}\to \frac 1g\tilde\psi_r$, we obtain the fermionic
kinetic term from \eqref{L0},
\begin{align}\label{fken}
{\rm tr}\big(i\Psi^{\dagger A}\gamma^\mu D_\mu\Psi_A\big)
= \frac1{g^2}{\rm tr}\big(i\tilde\psi_r\gamma^\mu \tilde D_\mu\tilde\psi_r\big),
\quad\quad r=1,\cdots,8.
\end{align}
Similarly the Higgsing of the ABJM fermionic
potential \eqref{Lfe} produces the following Yukawa type coupling
\begin{align}\label{Yukawa}
{\cal L}_{\rm Yukawa}=-\frac1{g^2} {\rm tr}\big\{\Gamma_{i}^{rs}\tilde\psi_r
[\tilde X^{i},\tilde\psi_s]\big\}, \qquad i=1,\cdots,7.
\end{align}
Here $\Gamma_i$$'$s are turned out to be the 7-dimensional Euclidean gamma
matrices satisfying
\begin{align}
\{\Gamma_i,\,\Gamma_j\} = -2\delta_{ij}.
\end{align}
In the current notation they are given by
\begin{align}\label{Gamma}
& \Gamma_1=-(i\sigma_2\otimes\Delta_2+\sigma_1\otimes\Delta_3),\quad
\Gamma_2=-(i\sigma_2\otimes\Delta_4+\sigma_1\otimes\Delta_5),\nonumber\\
&\Gamma_3=-(i\sigma_2\otimes\Delta_6+\sigma_1\otimes\Delta_7),\quad
\Gamma_4=i\sigma_2\otimes\Delta_1,\quad
\Gamma_5=-(\mathbb{I}\otimes\Delta_8-\sigma_2\otimes\Delta_3),\nonumber\\
&\Gamma_6=-(\mathbb{I}\otimes\Delta_9-\sigma_2\otimes\Delta_5),\quad
\Gamma_7=-(\mathbb{I}\otimes\Delta_{10}-\sigma_2\otimes\Delta_7),
\end{align}
where $\sigma_{1,2}$ are the first and the second Pauli matrices, while
$\Delta_k$ are $4\times4$ matrices,
\begin{align}
&\Delta_1^{pq}=\delta^{pq}-2 \delta^{p4}\delta^{q4},\quad
\Delta_2^{pq}= \delta^{p4}\delta^{q1}+\delta^{p1}\delta^{q4},\quad
\Delta_3^{pq}=\delta^{p2}\delta^{q3}-\delta^{p3}\delta^{q2},\nonumber\\
&\Delta_4^{pq}=\delta^{p4}\delta^{q2}+\delta^{p2}\delta^{q4},\quad
\Delta_5^{pq}=\delta^{p3}\delta^{q1}-\delta^{p1}\delta^{q3},\quad
 \Delta_6^{pq}=\delta^{p4}\delta^{q3}+\delta^{p3}\delta^{q4},\nonumber\\
&\Delta_7^{pq}=\delta^{p1}\delta^{q2}-\delta^{p2}\delta^{q1},\quad
\Delta_8^{pq}=\delta^{p4}\delta^{q1}-\delta^{p1}\delta^{q4},\quad
 \Delta_9^{pq}=\delta^{p4}\delta^{q2}-\delta^{p2}\delta^{q4},\nonumber\\
&\Delta_{10}^{pq}=\delta^{p4}\delta^{q3}-\delta^{p3}\delta^{q4}.
\end{align}

Collecting the results in (\ref{Lbos}),~(\ref{fken}), and (\ref{Yukawa}),
we obtain the Lagrangian of ($2+1$)-dimensional ${\cal N}=8$ SYM with U($N$) gauge
symmetry,
\begin{align}\label{N=8SYM}
{\cal L}_{\rm SYM}^{{\cal N}=8}
&=\frac 1{g^2}{\rm tr}\Big(-\frac12 \tilde F_{\mu\nu}\tilde F^{\mu\nu}
-\tilde D_\mu \tilde X^i \tilde D^\mu \tilde X^i
+ i\tilde\psi_r\gamma^\mu \tilde D_\mu\tilde\psi_r
+ \frac12[\tilde X^i,\,\tilde X^j]^2
-\Gamma_{i}^{rs}\tilde\psi_r [\tilde X^{i},\tilde\psi_s]\Big).
\end{align}
The supersymmetry variations of the gauge and the matter fields are given by
\begin{align}\label{susy-var}
&\delta_\epsilon
A^+_{\mu}=i\tilde\epsilon^{r}\gamma_{\mu}\tilde\psi_r,
\nonumber\\
&\delta_\epsilon
\tilde X^i=i\Gamma_i^{rs}\tilde\epsilon_r
\tilde\psi_s,
\nonumber\\
&\delta_\epsilon\tilde\psi_r=
i\tilde F_{\mu\nu}\sigma^{\mu\nu}\tilde\epsilon_r
+\Gamma_i^{rs}\gamma^{\mu}\tilde\epsilon_sD_\mu\tilde X^i
-\Gamma_{ij}^{rs}\tilde\epsilon_{s}[\tilde X^i,\tilde X^j],
\end{align}
where $\tilde\epsilon_r = \tilde\epsilon^r$ and
\begin{align}
\sigma^{\mu\nu}=-\frac i4\big(\gamma^\mu\gamma^\nu
-\gamma^\nu\gamma^\mu\big),\quad\quad \Gamma_{ij}=\frac
i4\big(\Gamma_i\Gamma_j-\Gamma_j\Gamma_i\big).
\end{align}
Six of these eight supersymmetries are inherited from the six
supersymmetries of the  ABJM theory. This can be easily identified by setting
$\tilde\epsilon_4= \tilde\epsilon_8=0$ in \eqref{susy-var}.
In addition to these six supersymmetries, additional two supersymmetries arise
as a consequence of the breaking of the gauge symmetry
in the MP Higgsing procedure which moves the M2-branes away from the orbifold singularity.

\section{Higgsing of WZ-type Coupling with Constant Flux}\label{fluxH}

For the ABJM theory, there is mass deformation which preserves
${\cal N}=6$ supersymmetry~\cite{Hosomichi:2008jb,Gomis:2008vc}. The
origin of this deformation is a WZ-type coupling to constant
transverse four-form field strength which is dual to constant
seven-form field strength. This coupling with a particular choice of
the constant field strength can be identified with the quartic self
interaction term of transverse scalars~\cite{KKNT}, while the
quadratic mass term is understood as a result of the backreaction of
the flux on the geometry~\cite{Lambert:2009qw}. In this section we
will consider more generic constant transverse four-form and the
dual seven-form field strengths and reduce the corresponding WZ-type
coupling to type IIA string theory via the Higgsing procedure.

First let us consider the WZ-type coupling for the three-form field $C_3$.
In the presence of a constant transverse four-form field strength $F_4$ the components of
the corresponding $C_3$  are independent of the worldvolume coordinates
and are at most linear in the transverse coordinates. More precisely, $C_{\mu\nu\rho},~
C_{\mu\nu A},~C_{\mu AB}$, and $C_{\mu A\bar B}$ are all constants while $C_{ABC}$ and
$C_{AB\bar C}$
are linear in the transverse coordinates.\footnote{We employ the same index notation as in
Ref.~\cite{KKNT}.} We set the constant components
to zero by gauge freedom of the three-form gauge field ($\delta C_3 = d\Lambda_2$).
Then the gauge invariant  WZ-type coupling for this particular choice of
the three-form gauge field can be read from the equation (2.3) of Ref.~\cite{KKNT},

\begin{align}\label{massC3}
S_{C}^{(3)} = \lambda\int d^3x\, &\frac{1}{3!}
\epsilon^{\mu\nu\rho}{\rm tr}\Big[ C_{A\bar B C} D_\mu Y^A D_\rho Y_B^\dagger D_\nu Y^C
+ ({\rm c.c.}) \Big],
\end{align}
where $\lambda=2\pi  l_{{\rm P}}^{3/2}$ with Planck length $l_{{\rm P}}$.

The dual seven-form field strength $F_7$ is given by
\begin{align}
F_7=\ast F_4+\frac12 C_3\wedge F_4.
\end{align}
This implies, in the presence of the constant transverse $F_4$ the
components of $F_{7}$ with all indices in the transverse directions
are linear in the scalar fields while the remaining components are
constants. Keeping this in mind, the gauge invariant WZ-type
coupling for the corresponding six-form gauge field $C_6$ can be
read from the equation (2.8) of Ref.~\cite{KKNT},
\begin{align}\label{act6}
S_{C}^{(6)} = -\frac{\pi}{k\lambda}\int &d^3x\, \frac{1}{3!}
\epsilon^{\mu\nu\rho} \left\{{\rm Tr}\right\}\Big[ C_{\mu\nu\rho A
B\bar C} \beta^{AB}_{~C}
+3\lambda^2 \big(C_{\mu ABC\bar D\bar E} D_\nu Y^A D_\rho Y_D^\dagger \beta^{BC}_{~E}\nonumber \\
&+C_{\mu AB\bar C\bar D\bar E} D_\nu Y_C^\dagger D_\rho
Y_D^\dagger \beta^{AB}_{~E} \big)
+\lambda^3\big(C_{ABCD\bar E\bar F}D_\mu Y^A D_\nu Y^B D_\rho Y_E^\dagger
\beta^{CD}_{~F} \\
&+C_{ABC\bar D\bar E\bar F}D_\mu Y^A D_\nu Y_D^\dagger D_\rho
Y_E^\dagger \beta^{BC}_{~F}
+C_{AB\bar C\bar D\bar E\bar F}D_\mu Y_C^\dagger D_\nu
Y_D^\dagger D_\rho Y_E^\dagger \beta^{AB}_{~F}\big) + ({\rm
c}.{\rm c}.) \Big],\nn
\end{align}
where $\beta^{AB}_{~C}\equiv\frac12(Y^AY_C^\dagger Y^B - Y^BY_C^\dagger Y^A)$ and
we have set constant components of $C_6$ to zero using gauge freedom.

Now we apply the MP Higgsing to the WZ-type couplings \eqref{massC3}
and \eqref{act6}. The three-form coupling in \eqref{massC3} and all
the six-form couplings in \eqref{act6} except for the first term
produce higher order in $\alpha^{'}$ after the Higgsing. Since we
are interested in the terms which are in the lowest order in
$\alpha^{'}$, we will neglect those higher terms and take into
account only the following WZ-type coupling
\begin{align}\label{massC6}
S_{C}^{(6)} = -\frac{\pi}{\lambda k}\int d^3x\, &\frac{1}{3!}
\epsilon^{\mu\nu\rho} {\rm tr}\big[ C_{\mu\nu\rho A
B\bar C} \beta^{AB}_{~C} + C^{\dagger}_{\mu\nu\rho AB\bar C}
(\beta_{~C}^{AB})^\dagger \big].
\end{align}
The six-form gauge fields which are linear in the transverse scalars are given by
\begin{align}\label{CC6}
C_{\mu\nu\rho A B\bar C} = -2\lambda
\epsilon_{\mu\nu\rho} T_{AB\bar C\bar D}Y_D^\dagger,\quad
C^\dagger_{\mu\nu\rho A B\bar C} = -2\lambda
\epsilon_{\mu\nu\rho} T_{AB\bar C\bar D}^*Y^D=-2\lambda
\epsilon_{\mu\nu\rho} T_{CD\bar A\bar B}Y^D,
\end{align}
where the complex-valued parameters
$T_{AB\bar C\bar D}$ are antisymmetric in the first two indices as well as
the last two indices.

The Higgsing procedure for the WZ-type coupling was established in more general setting
in Ref.~\cite{KKNT}. Along the same line the Higgsing of the \eqref{massC6} results in
the following Myers coupling,
\begin{align}\label{massC6-2}
\tilde S_{\tilde C}^{(5)} &= -\frac{i\pi v}{\lambda k} \int d^3x\, \frac{1}{3!}
\epsilon^{\mu\nu\rho} {\rm tr}(\tilde C_{\mu\nu\rho ij}[
\tilde X^i,\tilde X^j]), \quad i,j=1,\cdots, 7.
\end{align}
Here $\tilde X^i$'s are defined in \eqref{Xbar} and the R-R
five-form fields $\tilde C_{\mu\nu\rho ij}$ are identified as
\begin{align}
&\tilde C_{\mu\nu\rho ab}=-\frac i4\big(  C_{\mu\nu\rho a 4\bar b }
-  C_{\mu\nu\rho a 4\bar b}^\dagger -  C_{\mu\nu\rho b 4\bar a} +
C_{\mu\nu\rho b 4\bar a}^\dagger +  C_{\mu\nu\rho ab\bar4} -
C_{\mu\nu\rho ab\bar4}^\dagger \big),
\nonumber\\
&\tilde C_{\mu\nu\rho a4}=-\frac i2(  C_{\mu\nu\rho a4\bar4} -
C_{\mu\nu\rho a4 \bar4}^\dagger),\quad \tilde C_{\mu\nu\rho 4a+4}
=-\frac 12(  C_{\mu\nu\rho a4\bar4 }+  C_{\mu\nu\rho a 4\bar
4}^\dagger),
\nonumber\\
&\tilde C_{\mu\nu\rho ab+4}=-\frac 14\big(  C_{\mu\nu\rho a4\bar b}
+  C_{\mu\nu\rho a4\bar b}^\dagger +  C_{\mu\nu\rho b 4\bar a} +
C_{\mu\nu\rho b 4\bar a}^\dagger -   C_{\mu\nu\rho ab\bar4} -
C_{\mu\nu\rho ab\bar4}^\dagger \big),
\nonumber\\
&\tilde C_{\mu\nu\rho a+4b+4}= -\frac i4\big(  C_{\mu\nu\rho a4\bar
b} -  C_{\mu\nu\rho a4\bar b}^\dagger -  C_{\mu\nu\rho b 4\bar a} +
C_{\mu\nu\rho b 4\bar a}^\dagger -  C_{\mu\nu\rho ab\bar4} +
C_{\mu\nu\rho ab\bar4}^\dagger \big),\label{C5ij}
\end{align}
where $a,b = 1,2,3$. For the case of the linear six-form gauge field in
\eqref{CC6}, the corresponding R-R five-form gauge field is
\begin{align}\label{RR5}
\tilde C_{\mu\nu\rho ij} = -2\lambda\epsilon_{\mu\nu\rho}\tilde T_{ijk}\tilde X^k,
\end{align}
where $\tilde T_{ijk}$ are antisymmetric real-valued parameters.
Using \eqref{CC6}, \eqref{C5ij}, and \eqref{RR5} in the
action\eqref{massC6-2} and rescaling $\tilde X^i\, \to\,\frac1g
\tilde X^i$, we obtain Myers coupling for a constant R-R five-form
field in type IIA string theory~\cite{Myers:1999ps},
\begin{align}\label{XXX}
\tilde S_{\tilde C}^{(5)} = \frac{i}{g^2}\int d^3x\,
{\rm tr}\big(\tilde T_{ijk}\tilde X^i[\tilde X^j,\,\tilde X^k]\big)
\end{align}
with
\begin{align}\label{Uijk}
&\tilde T_{ab4} = \tilde T_{4a+4b+4}=-\frac{i}{2} \big(T_{a4\bar b \bar4}
-T_{b4\bar a \bar4}\big),\quad
\tilde T_{a4b+4} = \frac{1}{2} \big(T_{a4\bar b \bar4} +T_{b4\bar a \bar4}\big),
\nn \\
&\tilde T_{abc} =-\frac{i}{4}\big(T_{a4\bar b\bar c} + T_{c4\bar a\bar b}+T_{b4\bar c\bar a}
-T_{bc\bar a\bar 4} - T_{ab\bar c\bar 4}-T_{ca\bar b\bar 4}\big),
\nn \\
&\tilde T_{abc+4} =-\frac{1}{4}\big(T_{a4\bar b\bar c} - T_{c4\bar a\bar b}+T_{b4\bar c\bar a}
+T_{bc\bar a\bar 4} - T_{ab\bar c\bar 4}+T_{ca\bar b\bar 4}\big),
\nn \\
&\tilde T_{ab+4c+4} =\frac{i}{4}\big(T_{a4\bar b\bar c} - T_{c4\bar a\bar b}-T_{b4\bar c\bar a}
-T_{bc\bar a\bar 4} + T_{ab\bar c\bar 4}+T_{ca\bar b\bar 4}\big),
\nn \\
&\tilde T_{a+4b+4c+4} =-\frac{1}{4}\big(T_{a4\bar b\bar c} + T_{c4\bar a\bar b}+T_{b4\bar c\bar a}
+T_{bc\bar a\bar 4} + T_{ab\bar c\bar 4}+T_{ca\bar b\bar 4}\big).
\end{align}

In section \ref{massSYM}, we will consider the ${\cal N}=8$ SYM
theory discussed in section \ref{ABJM-H} and deform it by the Myers
coupling \eqref{XXX}. In general such deformation breaks the
supersymmetry. However, if one also include an appropriate quadratic
mass term and turn only some particular nonvanishing components of
$\tilde T_{ijk}$, it is possible to preserve some supersymmetries.

\section{Mass-deformations of (2+1)-dimensional SYM}\label{massSYM}

In section 2 we have seen that the Higgsing of the undeformed ABJM
theory led to the ${\cal N}=8$ SYM theory in (2+1)-dimensions, with
U($N$) gauge symmetry, however, the maximal supersymmetric ABJM
theory with the quadratic mass term does not have any flat direction
and as a result the MP Higgsing procedure cannot be applied to this
case. In ($3+1$)-dimensions some mass-deformed SYM theories have
already been constructed~\cite{Polchinski:2000uf}, where the origin
of deformations in ${\cal N}=1^*$ and ${\cal N}=2^*$ SYM theories
were interpreted as the Myers couplings of D3-branes with constant
background flux in type IIB string theory. We naturally expect that
dimensional reduction of these theories  results in
($2+1$)-dimensional mass-deformed SYM theories with certain amount
of supersymmetry.

In the framework of AdS/CFT correspondence, ($2+1$)-dimensional
mass-deformed SYM theories have been studied in
Refs.~\cite{Bena:2000zb,Bena:2000fz,Bena:2000va,Ahn:2001nw}. Along
the same line with the ($3+1$)-dimensional SYM
theories~\cite{Polchinski:2000uf}, mass term for the fermionic
fields was turned on in the ($2+1$)-dimensional field theory side
and the corresponding background flux in type IIA supergravity was
found. In the presence of the background flux, the D2-branes are
polarized into NS5-branes when all the fermions are massive or
polarized into D4-branes when one of the fermions is left massless.
Based on these brane interpretations, the M-theory origin of the
mass-deformed SYM theories was also discussed. The story and its
M-theory interpretation are not complete because the Lagrangian with
supersymmetry transformation rules was not explicitly written and
because the Lagrangian formulation of multiple M2-branes was not
known at the time. We will fill the gaps of the scenario in this and
the subsequent sections. Guided by supersymmetry invariance, we
obtain mass-deformations of ($2+1$)-dimensional SYM theories with
${\cal N}=1$, ${\cal N}=2$, and ${\cal N}=4$ supersymmetries. The
number of supersymmetries depends on the choice of the constant
background flux configuration generating the deformations in type
IIA string theory.

We start by considering a configuration of multiple D2-branes on a
background of constant transverse four-form field strength
$F_{ijkl}$ in the absence of NS-NS two-form field. Then the dual
six-form field strength is given by
\begin{align}
\frac1{6!}F_{\mu\nu\rho ijk}=\frac1{4!}\epsilon_{\mu\nu\rho
ijk}\!\!\!~^{i'j'k'l'}F_{i'j'k'l'},
\end{align}
and in the symmetric gauge
\begin{align}\label{C5c}
\tilde C_{\mu\nu\rho ij}= \tilde\lambda F_{\mu\nu\rho ijk}\tilde
X^k= \tilde\lambda\epsilon_{\mu\nu\rho}\tilde T_{ijk}\tilde X^k,
\end{align}
where $\tilde\lambda=2\pi l^2_{\rm s}$.
 The Myers coupling for this
five-form gauge field \eqref{C5c} gives cubic self-interaction terms
between the transverse scalar fields
\begin{align}\label{LXXX}
{\cal L}_{XXX}= \frac{i}{g^2}\, {\rm tr}\big(\tilde T_{ijk}\tilde
X^i[\tilde X^j,\,\tilde X^k]\big),
\end{align}
while the Myers coupling for the three-form gauge field is either a
constant term or higher $\alpha^{'}$ corrections, which we do not
take into account in the low energy effective theory of our
interest. By calculating the backreaction of the constant four-form
field strengths on the geometry, one may obtain the following
quadratic term
\begin{align}\label{LXX}
{\cal L}_{XX} = -\frac{1}{g^2}M_{ij}{\rm tr}\big(\tilde X^i\tilde
X^j\big),
\end{align}
where the specific form of the mass matrix $M_{ij}$ is controlled by
the supersymmetry of the mass-deformed theory.

\subsection{${\cal N}=1$ }

In this subsection we will determine a choice of the constant
six-form field strength which breaks all the supersymmetries of the
${\cal N}=8$ undeformed SYM but preserves ${\cal N}=1$. In this case
we can set all the supersymmetry parameters in \eqref{susy-var} to
zero except for one parameter. For instance we choose nonvanishing
$\tilde\epsilon_1$ and then set
$\tilde\epsilon_r=\epsilon\delta_{1r}$. Naturally we group the
fermionic fields as
\begin{align}
\tilde\lambda\equiv\tilde\psi_1, ~~{\rm
and}~~\tilde\xi_p\equiv\tilde\psi_p, ~{\rm for}~ p=2,\cdots,8.
\end{align}
Accordingly, the supersymmetry transformations \eqref{susy-var} of
the gauge and the matter fields are rewritten\footnote{In this
section the indices $p,q,\cdots$ run over the range $2,\cdots,8$.}
as
\begin{align}\label{N=1}
  &\delta
 A^+_{\mu}=i\epsilon\gamma_{\mu}\tilde\lambda,\quad\quad\delta
 \tilde X^i=i\Gamma_i^{1p}\epsilon
\tilde\xi_p,\quad\quad
\nonumber\\
&\delta\tilde\lambda=
i\tilde F_{\mu\nu}\sigma^{\mu\nu}\epsilon, \qquad
\delta\tilde\xi_p=
\Gamma_i^{p1}\gamma^{\mu}\epsilon D_\mu\tilde X^i
 -\Gamma_{ij}^{p1}\epsilon[\tilde X^i,\tilde X^j].
  \end{align}
The undeformed SYM action \eqref{N=8SYM} is manifestly invariant
under such reduced ${\cal N}=1$ supersymmetry.

In order to render the supersymmetry invariance of the deformed
theory, we introduce the following additional supersymmetry
transformation:
\begin{align}\label{N=1mass}
  &\delta'
 A^+_{\mu}=0,\quad\quad\delta'\tilde X^i= 0
 ,\quad\quad\delta'\tilde\lambda_a=0,\quad\quad\delta'\tilde\xi_{p}=
  \mu_{pq}\Gamma_i^{q1}\epsilon\tilde X^i,
\end{align}
where $\mu_{pq}=\mu_p\delta_{pq}$. In this setup the $\mu_p$'s are
mass parameters for the seven massive fermionic fields
$\tilde\xi_p$, which can have different values.
 When the fermionic mass term is chosen as
\begin{align}
{\cal L}^\mu_{\rm ferm} = -\frac{i}{g^2}\mu_{pq}\,{\rm
tr}(\tilde\xi_p\tilde\xi_q),
\end{align}
the total Lagrangian of our consideration
\begin{align}\label{N=1L}
{\cal L}^{{\cal N}=1}={\cal L}^{{\cal N}=8}_{\rm SYM}+{\cal L}_{
XX}+{\cal L}_{XXX}+{\cal L}^\mu_{\rm ferm},
\end{align}
becomes invariant under the total supersymmetry transformations in
\eqref{N=1} and \eqref{N=1mass}. This is possible if we choose
bosonic mass matrix $M_{ij}$ and the antisymmetric tensor $\tilde
T_{ijk}$ as
\begin{align}\label{MT}
&M_{ij}={\rm diag}(\mu_8^2,~\mu_7^2,~ \mu_6^2,~ \mu_5^2, ~\mu_4^2,~\mu_3^2,~\mu_2^2),
\nn \\
&\tilde T_{ijk}(\Gamma_k)^{1p} =\frac{1}3\mu_{qq'}\big(\Gamma_{i}^{1q}\Gamma_{j}^{q'p}-
   \Gamma_{j}^{1q}\Gamma_{i}^{q'p}\big)-\frac{2i}3\mu_{pq}\Gamma_{ij}^{1q}.
\end{align}
From the second line of \eqref{MT} we determine the nonvanishing
components of $\tilde T_{ijk}$:
\begin{align}\label{N1Tijk}
&\tilde T_{145}=-\frac 13(\mu_4+\mu_5+\mu_8),\quad
\tilde T_{246}=\frac 13(\mu_3+\mu_5+\mu_7),\quad
\tilde T_{347}=\frac 13(\mu_2+\mu_5+\mu_6),\nonumber\\
&\tilde T_{127}=-\frac 13(\mu_2+\mu_7+\mu_8),\quad
\tilde T_{136}=\frac 13(\mu_3+\mu_6+\mu_8),\quad
\tilde T_{235}=\frac 13(\mu_4+\mu_6+\mu_7),\nonumber\\
&\tilde T_{567}=-\frac 13(\mu_2+\mu_3+\mu_4).
\end{align}
The ${\cal N}=1$ mass-deformed SYM theory constructed here contains
one massless vector multiplet $(A_\mu^+,\,\tilde\lambda)$ and seven
massive matter multiplets $(\tilde X^i,\,\tilde\xi_p)$. As mentioned
previously the massive multiplets are allowed to have different
masses. \vskip 0.2cm

\noindent \underline{{\bf Fuzzy two ellipsoid solution:}}  From the
equations \eqref{N=8SYM}, \eqref{LXXX}, and \eqref{LXX} we read the
scalar potential of the ${\cal N}=1$ mass-deformed SYM theory
\eqref{N=1L}:
\begin{align}\label{Vscalar}
V(\tilde X)= -\frac{1}{g^2}\, {\rm tr}\Big(\frac12[\tilde X^i,\,\tilde
X^j]^2+i\tilde T_{ijk}\tilde X^i[\tilde X^j,\,\tilde X^k]-M_{ij}
\tilde X^i\tilde X^j\Big),
\end{align}
where the mass matrix $M_{ij}$ and the antisymmetric tensor $\tilde
T_{ijk}$ are given in \eqref{MT}--\eqref{N1Tijk}. The classical
supersymmetric vacuum equation satisfying $V(\tilde
X_{0})=\frac{\partial V(\tilde X_0)}{\partial \tilde X}=0$ can be
obtained from \eqref{N=1}--\eqref{N=1mass},
\begin{align}\label{N=1vac}
\Gamma_{ij}^{p1} [\tilde X^i,\tilde X^j]-
\mu_{pq}\Gamma_i^{q1}\tilde X^i =0\quad {\rm for}\,\, p=2,3,\cdots,
8.
\end{align}
The same equation can also be derived by component field expansion
of $F$-term equation in the ${\cal N} =1$ superfield formulation as
well. A nontrival solution to this vacuum equation is the
configuration of fuzzy two ellipsoid, which corresponds to D2-branes
polarized into
D4-branes~\cite{Bena:2000zb,Bena:2000fz,Bena:2000va,Ahn:2001nw}. The
fuzzy two ellipsoid can be embedded in any three-dimensional
transverse space with nonvanishing flux given by \eqref{N1Tijk} and
some constraints on the masses. Without loss of generality, here we
choose the directions $(1,4,5)$ as the embedding space of the fuzzy
two ellipsoid. The ansatz is
\begin{align}\label{susy-vac}
 \tilde X^i_0 = \left\{ \begin{array}{ll}
         \alpha_i T^{i} & \mbox{for\quad $i=1,4,5$},\quad ({\rm no~summation~over}~i)\\
        0 & \mbox{${\rm otherwise}$}\end{array} \right.
        \end{align}
where $T^i$'s are the generators of the $N$-dimensional irreducible
representation of SU(2) and $\alpha_i$ are real constants of
mass-dimension one. Then the $\tilde X^i_0$'s satisfy the
noncommutative algebra
\begin{align}\label{NC}[\tilde X^i_0,\,\tilde
X^j_0]=i\alpha_i\alpha_j\epsilon_{ijk}T^k ,\quad ({\rm
no~summation~over}~i,j)\end{align} which defines the fuzzy two
ellipsoid. Inserting \eqref{NC} into the vacuum equation
\eqref{N=1vac}, we obtain
 \begin{align}\label{vac-eq}
\alpha_1\alpha_4 -\alpha_5 \mu_4 =0,\qquad \alpha_1\alpha_5 -
\alpha_4\mu_5 =0,\qquad \alpha_4\alpha_5 -\alpha_1\mu_8 =0.
\end{align}
Assuming $\mu_4,\mu_5,\mu_8$ are either all positive or all
negative, we have the following set of solutions for \eqref{vac-eq}
\begin{align}
&\alpha_1= -\sqrt{\mu_4\mu_5},~\alpha_4=
\sqrt{\mu_4\mu_8},~\alpha_5= -\sqrt{\mu_5\mu_8},\nonumber\\
{\rm or}\quad&\alpha_1= -\sqrt{\mu_4\mu_5},~\alpha_4=
-\sqrt{\mu_4\mu_8},~\alpha_5= \sqrt{\mu_5\mu_8},\nonumber\\
{\rm or}\quad&\alpha_1= \sqrt{\mu_4\mu_5},~\alpha_4=
-\sqrt{\mu_4\mu_8},~\alpha_5= -\sqrt{\mu_5\mu_8},\nonumber\\
{\rm or}\quad&\alpha_1= \sqrt{\mu_4\mu_5},~\alpha_4=
\sqrt{\mu_4\mu_8},~\alpha_5= \sqrt{\mu_5\mu_8}.
\end{align}
The scale of the noncommutative space is characterized  by the size
of the semi-principal axes of the fuzzy two ellipsoid:
\begin{align}\label{radius}
R_i = 2\pi\alpha^{'}\sqrt{\frac 3N{\rm tr} [(\tilde X_0^i)^2]} =
\pi\alpha^{'}N\alpha_i\sqrt{1-\frac1{N^2}}\,,
\end{align}
where we have used ${\rm tr}(\tilde X_0^i)^2
=\frac1{12}\alpha_i^2N({N^2-1}).$ In the large $N$ limit,
\eqref{radius} is understood as the size of the shell D4-brane which
emerges as a result of polarization of D2-branes~\cite{Bena:2000fz}.
It is also important to note that this solution reduces to the fuzzy
two sphere solution if we started with $\alpha_1=\alpha_4=\alpha_5$.

\subsection{${\cal N}=2$ }

To find the ${\cal N}=2$ mass-deformed SYM theory we follow a
similar procedure as in the previous subsection. To that end we
start by assuming that two of the supersymmetric parameters in
\eqref{susy-var} are nonvanishing, for instance, $\tilde\epsilon_1$
and $\tilde\epsilon_2$. Then we can write the supersymmetry
parameters as $\tilde\epsilon_r=\epsilon_1\delta_{1r} +
\epsilon_2\delta_{2r}$ and group the fermionic fields as
\begin{align}
\tilde\lambda_a\equiv\tilde\psi_a, ~{\rm for}~ a=1,2 ~~{\rm
and}~~\tilde\xi_p\equiv\tilde\psi_p, ~{\rm for}~ p=3,4,\cdots,8.
\end{align}

Using the representation of seven-dimensional gamma matrices in
\eqref{Gamma}, we can write the supersymmetry transformations of the
gauge and the matter fields as\footnote{In this subsection we employ
the following indices. The bosonic field indices $m,n,\cdots$ are
used when the $i=7$ index is excluded, the fermionic field indices
$a,b,\cdots$ represent $1$ or $2$, the remaining fermionic field
indices $p,q,\cdots $ run over  $3,\cdots,8$.}
\begin{align}\label{N=2}
&\delta
 A^+_{\mu}=i\epsilon_{a}\gamma_{\mu}\tilde\lambda_a,\quad\quad\delta\tilde X^7=i\Gamma_7^{ab}\epsilon_a
\tilde\lambda_b,\quad\quad\delta\tilde X^m=i\Gamma_m^{ap}\epsilon_a\tilde\xi_p,\nonumber\\
 &\delta\tilde\lambda_a=
 i\tilde F_{\mu\nu}\sigma^{\mu\nu}\epsilon_a
 +\Gamma_7^{ab}\gamma^{\mu}\epsilon_bD_\mu\tilde X^7
 -\Gamma_{ij}^{ab}\epsilon_b[\tilde X^i,\tilde X^j],
 \nonumber\\
 &\delta\tilde\xi_p=
\Gamma_m^{pa}\gamma^{\mu}\epsilon_aD_\mu\tilde X^m
 -\Gamma_{ij}^{pa}\epsilon_a[\tilde X^i,\tilde X^j].
  \end{align}
Note that the undeformed SYM action \eqref{N=8SYM} is invariant
 under the reduced supersymmetry \eqref{N=2}.

 For the mass-deformed theory, along the same line with the ${\cal N}=1$
 SYM theory, we introduce the additional supersymmetry transformation:
\begin{align}\label{N=2mass}
&\delta' A^+_{\mu}=0,\quad\quad\delta'\tilde X^i= 0,
\quad\quad\delta'\tilde\lambda_a=0,\quad\quad\delta'\tilde\xi_{p}=
\mu_{pq} \Gamma_m^{qa}\epsilon_{a}\tilde X^m,
\end{align}
and a fermionic mass term
\begin{align}\label{ferm-mass}
{\cal L}^\mu_{\rm ferm} = -\frac{i}{g^2}\mu_{pq}{\rm
tr}\big(\tilde\xi_p\tilde\xi_q\big).
\end{align}
As in the case of ${\cal N}=1$, the mass matrix $\mu_{pq}$ is
diagonal, $\mu_{pq}=\mu_p\delta_{pq}$ and its elements $\mu_p$'s are
the mass parameters for the six massive fermionic fields, which are
not allowed to be all different.

From the invariance of the mass-deformed theory under
\eqref{N=2}--\eqref{N=2mass}, the mass matrix $M_{mn}$ of six
massive bosonic fields is determined as
\begin{align}
 M_{mn}={\rm diag}(\mu_8^2,~\mu_7^2,~ \mu_6^2,~ \mu_5^2, ~\mu_4^2,~\mu_3^2)
 \end{align}
with constraints
\begin{align}\label{bos-mass-rel} \mu_4^2=\mu_3^2,\quad~~ \mu_6^2=
\mu_5^2,\quad~~\mu_8^2=\mu_7^2.
\end{align}
The supersymmetry invariance also fixes the tensor $\tilde T_{ijk}$
as
\begin{align}
&\tilde T_{ijm}\Gamma_m^{ap}
=\frac{1}3\mu_{qq'}\big(\Gamma_{i}^{aq}\Gamma_{j}^{q'p}-
   \Gamma_{j}^{aq}\Gamma_{i}^{q'p}\big)-\frac{2i}3\mu_{pq}\Gamma_{ij}^{aq},
\nn\\
&   \tilde T_{ij7}\Gamma_7^{ab}
=\frac{1}3\mu_{pq}\big(\Gamma_{i}^{ap}\Gamma_{j}^{qb}-
   \Gamma_{j}^{ap}\Gamma_{i}^{qb}\big).
   \end{align}
From these equations we find the following nonvanishing components
of the tensor $\tilde T_{ijk}$,
\begin{align}
&\tilde T_{145}=\frac 13(\mu_3+\mu_6+\mu_7),\quad \tilde
T_{246}=\frac 13(\mu_3+\mu_5+\mu_7),\quad
\tilde T_{347}=\frac 13(\mu_5+\mu_6),\nonumber\\
&\tilde T_{127}=-\frac 13(\mu_7+\mu_8),\quad \tilde T_{136}=-\frac
13(\mu_4+\mu_5+\mu_7),\quad
\tilde T_{235}=-\frac 13(\mu_3+\mu_5+\mu_8),\nonumber\\
&\tilde T_{567}=-\frac 13(\mu_3+\mu_4),
\end{align}
where the $\mu_p$ should satisfy
\begin{align}\label{ferm-mass-rel}
\mu_3+\mu_4+\mu_5+\mu_6+\mu_7+\mu_8=0.
\end{align}

Here we have six massive fermionic fields $\tilde\xi_p$ and the same
number of massive bosonic fields $\tilde X^m$, which will form three
massive chiral multiplets of the ${\cal N}=2$ supersymmetry. In
addition we have a pair of massless fermionic fields
$\tilde\lambda_a$ and a single massless scalar $\tilde X^7$, which
together with the gauge boson $A_\mu^+$, form a massless vector
multiplet. \vskip 0.2cm

\noindent \underline{{\bf Fuzzy two ellipsoid solution:}} Along the
same line as ${\cal N}=1$ theory, the classical supersymmetric
vacuum equations are read from \eqref{N=2}--\eqref{N=2mass},
\begin{align}\label{N=2vac}
\Gamma_{ij}^{ab} [\tilde X^i,\tilde X^j]=0,\qquad\Gamma_{ij}^{pa}
[\tilde X^i,\tilde X^j]- \mu_{pq}\Gamma_i^{qa}\tilde X^i =0,\qquad
{\rm for}\,\, a,b=1,2,\,{\rm and}\,\, p=3,4,\cdots, 8.
\end{align}
The first equation is the component field expansion of $D$-term
equation, while the second is the expansion of the $F$-term equation
in ${\cal N} =2$ superfield formulation. The fuzzy two ellipsoid
solution to \eqref{N=2vac} is also obtained by using the ansatz
\eqref{susy-vac}. The $D$-term equation is trivially satisfied,
while the $F$-term equation leads to
 \begin{align}\label{vac-eq2}
&\alpha_1\alpha_4 + \alpha_5\mu_3 =0,\qquad \alpha_1\alpha_4 -
\alpha_5\mu_4 =0,\qquad
\alpha_1\alpha_5 -\alpha_4\mu_5 =0,\nn\\
&\alpha_1\alpha_5 + \alpha_4\mu_6 =0,\qquad \alpha_4\alpha_5 +
\alpha_1\mu_7 =0,\qquad \alpha_4\alpha_5 -\alpha_1\mu_8 =0.
\end{align}
In order to have nontrivial solution for $\alpha_i$'s, we should set
$\mu_4=-\mu_3,~\mu_6=-\mu_5,~\mu_8=-\mu_7$. Then, assuming
$\mu_3\mu_5<0,~\mu_3\mu_7>0,~{\rm and}~\mu_5\mu_7<0$, we obtain the
following set of solutions
\begin{align}
&\alpha_1 = -i\sqrt{\mu_3\mu_5},~\alpha_4 =
-\sqrt{\mu_3\mu_7},~\alpha_5 = -i\sqrt{\mu_5\mu_7},\nonumber\\
{\rm or}\quad&\alpha_1 = -i\sqrt{\mu_3\mu_5},~\alpha_4 =
\sqrt{\mu_3\mu_7},~\alpha_5 = i\sqrt{\mu_5\mu_7},\nonumber\\
{\rm or}\quad&\alpha_1 = i\sqrt{\mu_3\mu_5},~\alpha_4 =
\sqrt{\mu_3\mu_7},~\alpha_5 = -i\sqrt{\mu_5\mu_7},\nonumber\\
{\rm or}\quad&\alpha_1 = i\sqrt{\mu_3\mu_5},~\alpha_4 =
-\sqrt{\mu_3\mu_7},~\alpha_5 = -i\sqrt{\mu_5\mu_7}.
\end{align}
The size of the fuzzy two ellipsoid and the D-brane interpretation
of these solutions are the same as those of the ${\cal N}=1$ case in
\eqref{radius}.

\subsection{${\cal N}=4$}

Finally, we will find the choice of the constant flux which
preserves ${\cal N}=4$ supersymmetry. The detail is as in the
pervious subsection. The supersymmetry transformations of the gauge
and the matter fields are\footnote{The index notation for this
subsection is: the massive bosonic field indices $m,n,\cdots$ run
over the range $1,\cdots,4$, while the massless bosonic field
indices $\tilde m,\tilde n,\cdots$ are $5,6,7$; the massless
fermionic field indices $a,b,\cdots$ run over the range
$1,\cdots,4$; the massive fermionic field indices $p,q,\cdots$ run
over the range $5,\cdots,8$.}
\begin{align}
  &\delta
 A^+_{\mu}=i\epsilon_{a}\gamma_{\mu}\tilde\lambda_a,\quad\quad
 \delta\tilde X^{\tilde m}=i\Gamma_{\tilde m}^{ab}\epsilon_a
\tilde\lambda_b,\quad\quad\delta\tilde X^m=i\Gamma_m^{ap}\epsilon_a\tilde\xi_p,\nonumber\\
 &\delta\tilde\lambda_a=
 i\tilde F_{\mu\nu}\sigma^{\mu\nu}\epsilon_a
 +\Gamma_{\tilde m}^{ab}\gamma^{\mu}\epsilon_bD_\mu\tilde X^{\tilde m}
 -\Gamma_{ij}^{ab}\epsilon_b[\tilde X^i,\tilde X^j],
\nonumber\\
 &\delta\tilde\xi_p=
\Gamma_m^{pa}\gamma^{\mu}\epsilon_aD_\mu\tilde X^m
 -\Gamma_{ij}^{pa}\epsilon_a[\tilde X^i,\tilde X^j].
  \end{align}
The additional supersymmetry transformation and the fermionic mass
term are given by \eqref{N=2mass}--\eqref{ferm-mass} with the
indices adjusted to the notation of this subsection. From the
invariance of the mass-deformed theory we have the mass matrix
$M_{mn}$ of four massive bosonic fields,
\begin{align}\label{Mmn}
 M_{mn}=\mu^2 {\rm diag}(1,1,1,1), \quad
{\rm with}\,\,\,\mu^2=\mu_5^2=\mu_6^2= \mu_7^2=\mu_8^2.
 \end{align}

As usual, supersymmetry invariance determine the following relations
for $\tilde T_{ijk}$
\begin{align}
&\tilde T_{ijm}\Gamma_m^{ap}
=\frac{1}3\mu_{qq'}\big(\Gamma_{i}^{aq}\Gamma_{j}^{q'p}-
   \Gamma_{j}^{aq}\Gamma_{i}^{q'p}\big)-\frac{2i}3\mu_{pq}\Gamma_{ij}^{aq},
\nn\\
&\tilde T_{ij\tilde m}\Gamma_{\tilde m}^{ab}
=\frac{1}3\mu_{pq}\big(\Gamma_{i}^{ap}\Gamma_{j}^{qb}-
   \Gamma_{j}^{ap}\Gamma_{i}^{qb}\big).
   \end{align}
These equations leave the nonvanishing components
\begin{align}
&\tilde T_{145}=\frac 13(\mu_6+\mu_7),\quad \tilde T_{246}=\frac
13(\mu_5+\mu_7),\quad
\tilde T_{347}=\frac 13(\mu_5+\mu_6),\nonumber\\
&\tilde T_{127}=-\frac 13(\mu_7+\mu_8),\quad \tilde T_{136}=-\frac
13(\mu_5+\mu_7),\quad \tilde T_{235}=-\frac 13(\mu_5+\mu_8),
\end{align}
where the $\mu_p$'s should satisfy
\begin{align}\label{ferm-mass-rel2}
\mu_5+\mu_6+\mu_7+\mu_8=0.
\end{align}

Now we have one massless vector multiplet $(A_\mu^+,\,\tilde X^m,\,
\tilde \lambda_a)$ and one massive chiral multiplet $(\tilde
X^{\tilde m},\, \tilde\xi_p)$. It is also important to note that the
relations \eqref{Mmn} and \eqref{ferm-mass-rel2} for the mass matrix
leads to unique choice for the fermionic mass matrix $\mu_{pq}=\mu\,
{\rm diag}(1,1,-1,-1)$ since the other choices are equivalent to
this one up to field redefinitions. \vskip 0.2cm

\noindent \underline{{\bf Nonexistence of fuzzy two ellipsoid
solution:}} Substituting the ansatz \eqref{susy-vac} into the vacuum
equation \eqref{N=2vac}, we rewrite the $D$- and $F$-term vacuum
equations for ${\cal N}=4$ SYM theory as
 \begin{align}\label{vac-eq3}
&\alpha_1\alpha_4  =0,\quad\alpha_1\alpha_5 -\alpha_4\mu_5 =0,\quad
\alpha_1\alpha_5 + \alpha_4\mu_6 =0,\quad \alpha_1\alpha_5 +\alpha_4
\mu_7 =0,\quad \alpha_1\alpha_5 -\alpha_4\mu_8
=0,\nonumber\\
&\alpha_4\alpha_5 -\alpha_1\mu_5 =0,\quad \alpha_4\alpha_5 +
\alpha_1\mu_6 =0,\quad \alpha_4\alpha_5 +\alpha_1\mu_7 =0,\quad
\alpha_4\alpha_5 -\alpha_8\mu_8 =0.
\end{align}
In this case, one easily notices that there is only a trivial
solution for which
 $\alpha_1=\alpha_4=0$ and $\alpha_5$ is any real number. Thus the
nontrivial fuzzy two ellipsoid configuration can not be a classical
supersymmetric vacuum solution in ${\cal N}=4$ theory as expected.

\section{M-theory Origin of Mass-deformed SYM}\label{Morg}

In this section we identify the M-theory origin of the flux terms
for each of the mass-deformed SYM theories discussed in the previous
section. This can be achieved by comparing the supersymmetry
preserving flux backgrounds of the previous section to the results
of the Higgsing procedure of section \ref{fluxH}. We invert the
relations in \eqref{Uijk} and find the constant flux in M-theory in
terms of the antisymmetric parameters $\tilde T_{ijk}$ \eqref{RR5}
in type IIA string theory. To be specific this results in the
following
 relations for $T_{AB\bar C\bar D}$ of \eqref{CC6},
\begin{align}\label{Mflux}
&T_{a4\bar b\bar 4} = i\tilde T_{ab4} + \tilde T_{a4b+4}, \quad {\rm
with}\,\,\,\tilde T_{a4b+4}=\tilde T_{b4a+4},
\nn \\
&T_{a4\bar b\bar c} = \tilde T_{bca+4} - \tilde T_{a+4b+4c+4} + i \big(
\tilde T_{abc} - \tilde T_{ab+4c+4}\big),
\end{align}
where all the other components vanish.

The nonvanishing components of $\tilde T_{ijk}$ take different
values depending on the number of supersymmetries. Here we list the
nonvanishing components of $T_{AB\bar C\bar D}$ for the three cases
separately. For the case of ${\cal N}=1$ supersymmetry, they are
\begin{align}
&T_{14\bar 1\bar 4} = \tilde T_{145}=-\frac 13(\mu_4+\mu_5+\mu_8),
\quad T_{24\bar 2\bar 4} =T_{246}=\frac 13(\mu_3+\mu_5+\mu_7),
\nn \\
&T_{34\bar 3\bar 4} =\tilde T_{347}=\frac 13(\mu_2+\mu_5+\mu_6),
\quad T_{14\bar 2\bar 3} =\tilde T_{235} - \tilde T_{567}
= \frac13 \big(\mu_2 + \mu_3 + 2\mu_4 + \mu_6 + \mu_7\big),
\nn \\
&T_{24\bar 1\bar 3} =\tilde T_{136} + \tilde T_{567}
= \frac13 \big(-\mu_2 -\mu_4 + \mu_6 + \mu_8\big),
\nn \\
&T_{34\bar 1\bar 2} =\tilde T_{127} - \tilde T_{567} = \frac13
\big(\mu_3 + \mu_4 - \mu_7 - \mu_8\big).
\end{align}
For the case of ${\cal N}=2$ supersymmetry, they are
\begin{align}
&T_{14\bar 1\bar 4} = \tilde T_{145}=-\frac 13(\mu_4+\mu_5+\mu_8),
\quad T_{24\bar 2\bar 4} =T_{246}=\frac 13(\mu_3+\mu_5+\mu_7),
\nn \\
&T_{34\bar 3\bar 4} =\tilde T_{347}=\frac 13(\mu_5+\mu_6),
\quad T_{14\bar 2\bar 3} =\tilde T_{235} - \tilde T_{567}
= \frac13 \big( \mu_4 -\mu_5 - \mu_8 \big),
\nn \\
&T_{24\bar 1\bar 3} =\tilde T_{136} + \tilde T_{567}
= \frac13 \big( -\mu_4 + \mu_6 + \mu_8\big),
\nn \\
&T_{34\bar 1\bar 2} =\tilde T_{127} - \tilde T_{567}
= \frac13 \big(\mu_3 + \mu_4 - \mu_7 - \mu_8\big),
\end{align}
For the case of ${\cal N}=1$ supersymmetry, they are
\begin{align}
&T_{14\bar 1\bar 4} = \tilde T_{145}=-\frac 13(\mu_5+\mu_8),
\quad T_{24\bar 2\bar 4} =T_{246}=\frac 13(\mu_5+\mu_7),
\nn \\
&T_{34\bar 3\bar 4} =\tilde T_{347}=\frac 13(\mu_5+\mu_6),
\quad T_{14\bar 2\bar 3} =\tilde T_{235}
= -\frac13 \big( \mu_5 + \mu_8 \big),
\nn \\
&T_{24\bar 1\bar 3} =\tilde T_{136}
= \frac13 \big( \mu_6 + \mu_8\big),
\quad T_{34\bar 1\bar 2} =\tilde T_{127}
= -\frac13 \big(\mu_7 + \mu_8\big).
\end{align}

The constant flux \eqref{Mflux} is determined in M-theory and is
different from the maximal supersymmetry preserving flux in ABJM
theory~\cite{Hosomichi:2008jb,Gomis:2008vc}. Therefore, with this
constant flux the maximal ${\cal N}=6$ supersymmetry is not
preserved in the ABJM theory. When such flux is turned on in
M-theory and the masses of the fermionic and bosonic fields are
appropriately chosen, the supersymmetry is partially preserved. In
these cases some of the scalar fields remain massless and the
corresponding transverse directions may become flat. Unlike the ABJM
theory with maximal supersymmetry
 preserving mass-deformation,
the MP Higgsing procedure can be applied to these theories
preserving partial supersymmetries. We naturally expect the end
result of such procedure will be the mass deformed SYM theories
discussed in section \ref{massSYM}.

\section{Conclusion}\label{conc}
In this paper we found supersymmetry preserving mass-deformations of
the SYM theory in (2+1)-dimensions and identified their M-theory
origin. In achieving this goal, we followed the current trend of
deriving supersymmetric Yang-Mills matter theories from the
Chern-Simons matter theories via the MP Higgsing procedure. In
particular, we considered the ABJM theory in the background of
arbitrary constant four-form field strength in the directions
transverse to the M2-branes. Actually, for our purpose of generating
supersymmetry preserving mass deformations of SYM theories, the
WZ-type coupling to the dual seven-form strength is important, while
the Higgsing of the corresponding coupling to the constant
transverse four-form strength gives either a constant term or higher
order $\alpha'$ corrections.

It has already been confirmed that a particular constant transverse
four-field strength and the dual-seven form field strength led to
the maximal supersymmetry preserving mass-deformation in the context
of the ABJM theory. In this paper we show that the MP Higgsing of
the original ABJM theory without mass-deformation reduces to the
(2+1)-dimensional ${\cal N}=8$ SYM theory without any extra
decoupled sector in contrast to the claim of an earlier work on the
same subject \cite{Pang:2008hw}. The result suggests that the
corresponding reduction of the maximal supersymmetric mass-deformed
ABJM theory might give some known mass deformed SYM theory. This
naive expectation did not work because in this case the scalar
potential does not contain flat direction and then one could not
take the
 limit of large vacuum expectation value in pursuing the Higgsing procedure.
  We instead began with the ABJM theory deformed by a WZ-type coupling to a
  constant
seven-form field strength which is dual to an arbitrary constant
four-form field strength in the transverse direction. This
deformation breaks supersymmetry but leaves us with some flat
directions. Application of the MP Higgsing procedure to the ABJM
theory deformed by this WZ-type coupling led to Yang-Mills matter
theories including a Myers coupling to a five-form gauge field with
constant six-form field strength. By different choices of the
nonvanishing components of the five-form gauge field and appropriate
identification of the corresponding masses for the fermionic and
bosonic fields, we obtained SYM theories with ${\cal N}=1$, ${\cal
N}=2$, and ${\cal N}=4$ supersymmetries. We solved the vacuum
equations for each of these theories and found the fuzzy two
ellipsoid solutions in the first two cases while in the third case
the equations support only the trivial solution. The obtained fuzzy
two ellipsoid solutions confirm that the D2-branes system polarizes
into a D4-branes system with the extra dimensions warping the two
ellipsoid when we turn on the mass terms for the matter fields
~\cite{Bena:2000fz}.

Finally, we used the values of the nonvanishing components of the
R-R five-form and determined the corresponding six-form gauge field
in M-theory. This may identify a possible M-theory origin of the
supersymmetry preserving mass-deformations of the SYM theories. It
is interesting to employ the supersymmetry completion to find the
appropriate quadratic mass-deformations in the ABJM theory and
figure out which of the supersymmetries of the theory are preserved
despite of such deformation. One can then apply the MP Higgsing
procedure to the reduced supersymmetric theories and expect to
reproduce the mass-deformed SYM theories we obtained in this paper.
These points and the issue concerning the dual
gravity~\cite{Kim:2010mr,Cheon:2011gv,Hashimoto:2011nn} of the
reduced supersymmetric theories will be reported in a separate
work~\cite{KKT}.

\section*{Acknowledgements}
This work was supported by the Korea Research Foundation Grant
funded by the Korean Government  with grant numbers
KRF-2008-313-C00170 and 2011-0011660 (Y.K.), 2009-0073775,
2011-0009972 (O.K.), and 2009-0077423 (D.D.T.).

\end{document}